\documentclass[fleqn,10pt]{wlscirep}

\title{Work and information from thermal states after subtraction of energy quanta}

\author[1]{J. Hlou\v{s}ek}

\author[1]{M. Je\v{z}ek}

\author[1,*]{R. Filip}

\affil[1]{Department of Optics, Palack\' y University, 17. listopadu 1192/12,  771~46 Olomouc,  Czech Republic}

\affil[*]{filip@optics.upol.cz}

\begin{abstract}
Quantum oscillators prepared out of thermal equilibrium can be used to produce work and transmit information. By intensive cooling of a single oscillator, its thermal energy deterministically dissipates to a colder environment, and the oscillator substantially reduces its entropy. This out-of-equilibrium state allows us to obtain work and to carry information. Here, we propose and experimentally demonstrate an advanced approach, conditionally preparing more efficient out-of-equilibrium states only by a weak dissipation, an inefficient quantum measurement of the dissipated thermal energy, and subsequent triggering of that states. Although it conditionally subtracts the energy quanta from the oscillator, average energy grows, and second-order correlation function approaches unity as by coherent external driving. On the other hand, the Fano factor remains constant and the entropy of the subtracted state increases, which raise doubts about a possible application of this approach. To resolve it, we predict and experimentally verify that both available work and transmitted information can be conditionally higher in this case than by arbitrary cooling or adequate thermal heating up to the same average energy. It qualifies the conditional procedure as a useful source for experiments in quantum information and thermodynamics.

\end{abstract}

\begin{document}

\flushbottom

\maketitle

\thispagestyle{empty}

\section*{Introduction}

Matter and radiation out of thermal equilibrium with an environment are significant resources of modern physics, information science, and technology. Thermal state of a cooled system is not in thermal equilibrium with its environment and can be used to perform work \cite{Greiner} and carry information \cite{Helstrom}. Preparation of the cooled state requires only a connection to a cold external reservoir where a large part of energy dissipates and, simultaneously, entropy gradually decreases.
Similarly, by thermal heating from an {\em external stochastic} hot reservoir, we can enlarge mean energy,  
but also the entropy. A high-energy out-of-equilibrium state can also be prepared by {\em external deterministic} force \cite{Glauber} applied to a thermal state. Such coherent driving renders the entropy lower than of the initial thermal state.
It is the best classical way for the preparation of states capable of transmitting more information \cite{Helstrom} and producing more work \cite{Gelbwaser,Lutz,Kiesel,Mari,Kolar,Paternostro}. Alternatively, mechanisms that do not require either external heating or driving allow us to test non-equilibrium quantum thermodynamics merging with information theory \cite{rev1,rev2}, also at currently unexplored experimental platforms.

Quantum optics has proven to be a suitable experimental platform for proof-of-principle tests of many quantum physics processes, heavily stimulating other experimental platforms and advancing novel quantum technologies. A weak dissipation of thermal energy of light to cold reservoir modes allow us to conditionally subtract individual quanta of that thermal energy by measuring the reservoir modes by quantum detectors. The subtraction can be successfully performed even under imperfect conditions and by using inefficient photodetectors.
In quantum optics, such photodetection processes continuous in time were first discovered to conditionally manipulate the statistics and also increase the energy of thermal light \cite{reff1,reff2,reff3,reff4}. Over two decades, the continuous-time nonclassical state manipulations were extensively experimentally developed in cavity quantum electrodynamics \cite{reff5, reff6}. During the same period, the series of multi-photon subtraction experiments with thermal light also demonstrated a conditional instantaneous increase of mean energy by a subtraction of quanta from single-mode thermal state \cite{sub1,sub2,sub3}.
The subtraction procedure use a basic dissipation mechanism, with no other energy supply, and an inefficient measurement of energy quanta without their exact resolution.
This procedure applied to classical states has been used for quantum filtering \cite{app1}, state preparation \cite{app2}, noiseless amplification \cite{app4}, quantum cloning \cite{app5}, enhanced interferometry \cite{app6,app7} and recently, also to illustrate Maxwell demon in quantum thermodynamics \cite{app8}. Different measures have been applied to quantify the effect of subtraction procedures \cite{quan1,quan2,quan3,quan4}. However, no analysis, experiment, or operational measures proving the principal applicability the subtracted thermal states in the information transmission and work extraction have not yet been presented.

Here, we experimentally verify that the instantaneous subtraction of a number of quanta (photons) from the thermal energy of the oscillator produces out-of-equilibrium state with increased average energy, but simultaneously it keeps Fano factor constant \cite{Fano}. It means that average energy increases hand-in-hand with its variance. Also, entropy slowly increases with the increasing number of subtracted quanta. Despite this limitation, we predict and demonstrate that such out-of-equilibrium states can provide work and carry information larger than what is available by any dissipative cooling mechanism. Photons-subtracted thermal states represent a paramount example of out-of-equilibrium states that can be obtained without an external coherent deterministic drive or an additional thermal source of energy. These states can be employed as a useful source for the various future experiments in currently joining fields of information theory and nonequilibrium quantum thermodynamics.

\begin{figure}
\centerline{\includegraphics[width=0.82\linewidth]{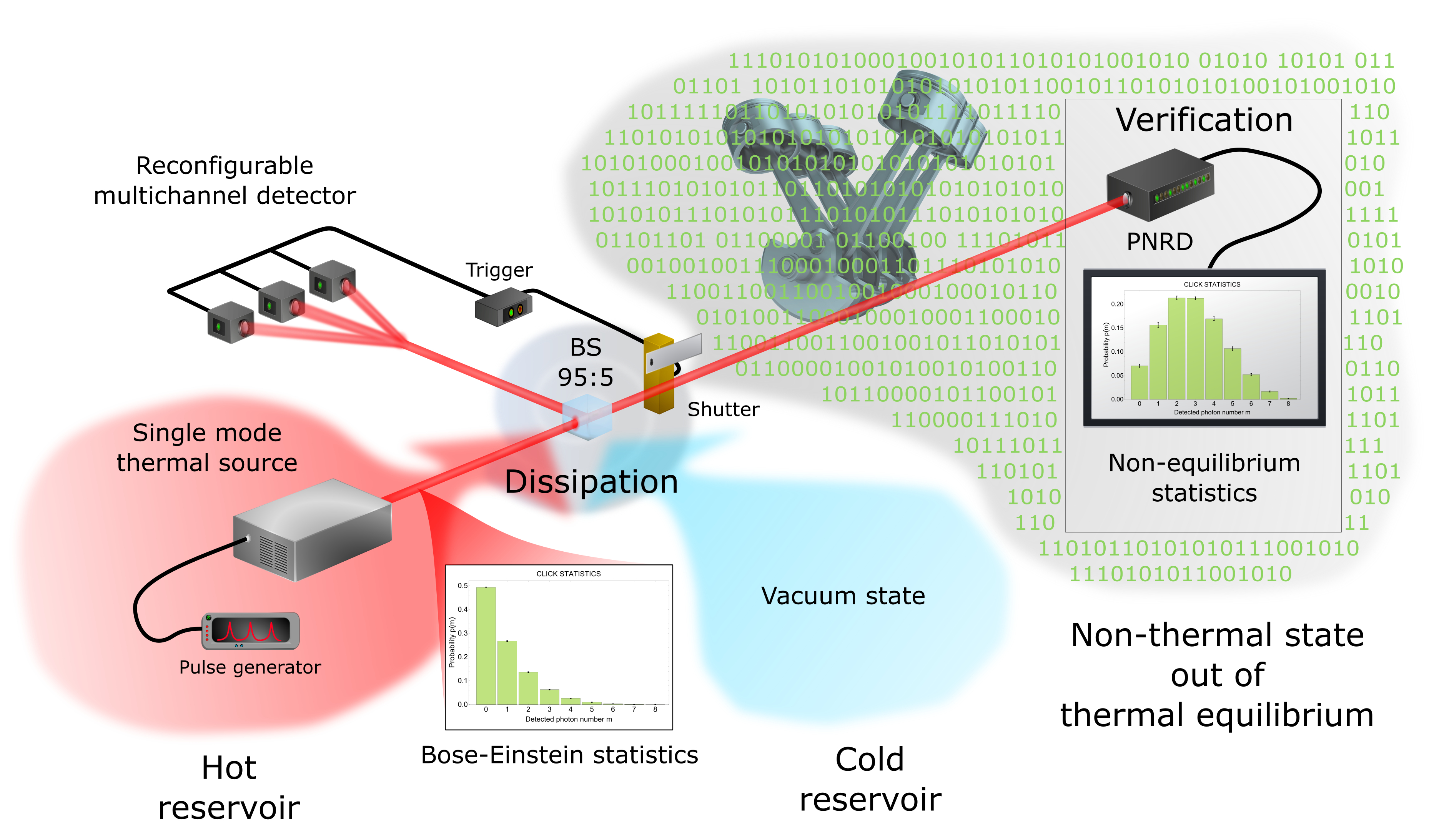}}
\caption{Preparation and characterization of out-of-equilibrium states conditionally generated via multiple-photon subtraction from single-mode thermal light. Thermal light governed by Bose-Einstein statistics dissipates at an unbalanced beam splitter (BS) to the vacuum reservoir modes. A small fraction of light in the reservoir modes is detected by a multichannel detector formed by $m$ on-off detectors. Coincidence detection events, when all the $m$ detectors fires, trigger the output and verification stage consisting of a photon-number-resolving detector. Subsequently, data are processed, and photon statistics of the conditionally prepared out-of-equilibrium state is analyzed. The statistics is evaluated for its capability to provide work and carry information.}
\label{fig:scheme}
\end{figure}

\section*{Subtraction of energy from thermal state}

To experimentally produce and analyze the out-of-equilibrium state of a {\em single} oscillator and demonstrate its capabilities we follow a stream of the optical experiments \cite{sub1,sub2,sub3,app1,app2,app4,app5,app6,app7,app8}. Our motivation and evaluation are however different. The scheme is depicted in Fig.~\ref{fig:scheme}.
The generation starts from a single oscillator represented by a single mode of radiation prepared in the state $\rho_{\text{th}}=\sum_{n=0}^{\infty}p_{n,\text{th}}|n\rangle\langle n|$ with thermal Bose-Einstein statistics $p_{n,\text{th}}=\frac{n_{\text{th}}^n}{\left(1+n_{\text{th}}\right)^{1+n}}$ determined only by the mean number $n_{\text{th}}$ of energy quanta, where $|n\rangle$ are energy basis states. In our experiment, thermal light is generated by temporal intensity modulation of a pulsed laser by rotating ground glass. The thermal state instantaneously dissipates small part of its energy at the unbalanced beam splitter to a multimode reservoir $R$ in vacuum (ground) state $|0\rangle_R$. The dissipation only negligibly cools down the thermal mode. Chiefly, it correlates states $|n\rangle$ of the oscillator's mode with a global photon number state $|k\rangle_R$ of the reservoir. It is apparent from the transformation
\begin{equation}\label{eq1}
|n\rangle\langle n|\otimes|0\rangle_R\langle 0| \rightarrow \sum_{k=0}^n {n\choose k} p_s^k (1-p_s)^{n-k}|n-k\rangle\langle n-k|\otimes |k\rangle_R\langle k|,
\end{equation}
where $p_s$ is a survival probability of single quantum in the oscillator. High single quantum survival probability $p_s$ means weak coupling. Product $p_s^k(1-p_s)^{n-k}$ stands for the probability that $k$ quanta will remain in the oscillator and $n-k$ quanta go to the reservoir $R$. In the experiment, only 5\% of the energy is dissipated, so $p_s=0.95$.

This entirely classical correlation between the system and reservoir at a level of individual quanta is a useful resource. It arises from the classical (first-order) coherence of single-mode thermal light \cite{Glauber}. For heavily multimode thermal oscillator (incoherent), the statistics of quanta over all weakly occupied modes approaches Poissonian, and the dissipative process does not produce this correlation. It means we cannot modify statistics by any measurement performed on the reservoir $R$. The multimode thermal light establishes an incoherent (classical) limit. To realize the importance of first-order coherence for the formation of the out-of-equilibrium state, the subtraction experiment with multimode thermal states is also performed. $M$ temporal thermal modes with the same overall mean photon number $\langle n\rangle=n_{\text{th}}$ are selected to prepare a multimode state. The effective number of modes $M$ is modified by changing the size of speckle pattern collected after the ground glass. The partially coherent $M$-mode state would produce an interference visibility of $1/M$ given by first-order coherence function $g^{1}(0)$.

Light dissipated to reservoir further scatters to many modes. To detect at least a small fraction of the dissipated light, we select $m$ modes and detect them by single-photon avalanche diodes (SPADs). Only when $m$-fold coincidence is detected, the resulting optical output of the source is transmitted. The ideal version of this detection can be described by \cite{Sperling12}
\begin{equation}\label{eq2}
\Pi_{m}={N \choose m}\sum_{s=m}^{\infty}
\sum_{j=0}^m\frac{1}{N^s}{m \choose j}(-1)^j(m-j)^s|s\rangle_A\langle s|,
\end{equation}
however, a real measurement collects only a small part of overall thermal energy dissipated into the reservoir $R$. Therefore, we introduce an overall effective collection efficiency $\eta$ by the transformation $|k\rangle\langle k| \rightarrow \sum_{r=0}^k {k\choose r} \eta^{k-r} (1-\eta)^{r}|k-r\rangle\langle k-r|$ of the energy states before the detection. Eqs.~(\ref{eq1},\ref{eq2}), together with the collection efficiency $\eta$, completely describe the instantaneous multiphoton subtraction process.

For a weak dissipative coupling with sufficiently high single-photon survival probability, $p_s\approx 1$, the out-of-equilibrium statistics approaches
\begin{equation}\label{subtr}
p_n=\frac{\frac{(n+m)!}{n!m!}\left(\frac{n_{\text{th}}}{1+ n_{\text{th}}}\right)^n}{(1+n_{\text{th}})^{m+1}}
\end{equation}
by conditioning on $m$ detection events. A potentially small $\eta\ll 1$ reduces the generation rate, but the prepared out-of-equilibrium states are very close to the theoretical limit. Importantly, (\ref{subtr}) describes a single-mode light. Its statistics is purely mathematically analogical to the overall statistics of $m+1$-mode thermal light equally populated in all the modes by an average number $n_{\text{th}}$ of quanta \cite{MR}.
In the case of multimode light, the different $m+1$ modes are principally distinguishable, and the light possesses lower first-order coherence quantified by $g^{1}(0)=1/(m+1)$. Also, available energy, work, and information {\em per mode} are actually $m+1$ times lower, because the distinguishable modes are not used efficiently. Consequently, the single-mode state with the statistics (\ref{subtr}) produced by a coherent light source thermodynamically outperforms multimode states with the same statistics and is better suited to our purpose.

In the verification stage of the experiment, the generated out-of-equilibrium statistics is independently analyzed by a photon-number-resolving detector (PNRD). The verification PNRD consists of tunable free-space multichannel optical network and sufficient number of SPADs, eight in our case, and features precise balancing with no crosstalk between the individual detection ports. Further experimental details and characterization of optical set-up are presented in the Methods section.

\section*{Out-of-equilibrium statistics}

\begin{figure}
\centerline{\includegraphics[width=0.7\textwidth]{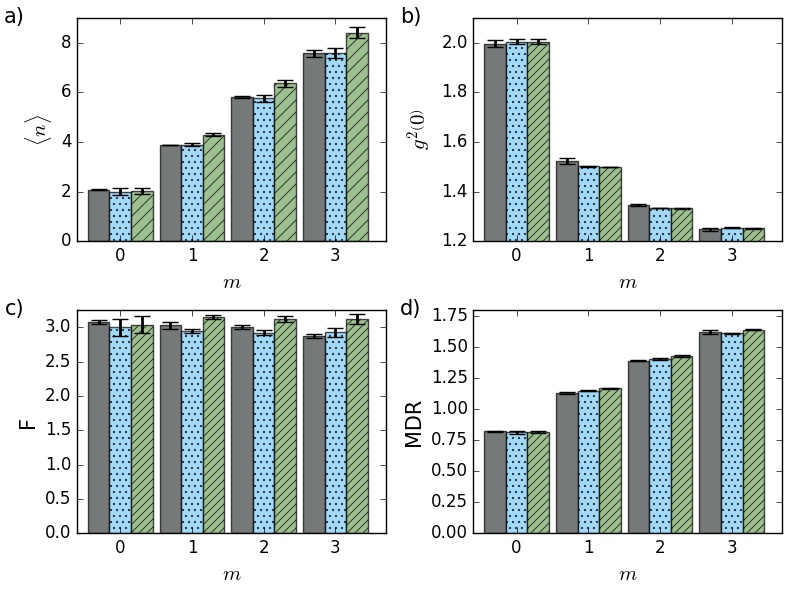}}
\caption{Mean number of photons (a), $g^{(2)}(0)$ function (b), Fano factor (c), and MDR (d) of the equilibrium ($m=0$) and out-of-equilibrium ($m>0$) states. $m$ stands for a number of subtracted photons. Experimental results (dark gray), full numerical model (blue dots), and the simplified model (\ref{subtr}) (green tiles) -- see the Methods for details on the theoretical models.
Data error bars show the standard deviation of the measurement, error bars of the models represent the uncertainty of input parameters, particularly of the mean number of photons determined from the measured initial thermal statistics.}
\label{fig:prms}
\end{figure}

We will analyze several essential parameters of the prepared out-of-equilibrium light governed by the statistics (\ref{subtr}) to assess its performance. As has been already discussed, the single-mode statistics (\ref{subtr}) yields a linearly increasing mean number of quanta, $\langle n\rangle=(m+1)n_{\text{th}}$, where $m$ in a number of conditional detection events. The monotonous increase is shown in Fig.~\ref{fig:prms}(a) for $n_{\text{th}}=2$ set in our measurement. It is important to stress that the behavior of the mean number of quanta of the state subjected to the subtraction process depends on the initial state statistics. The mean energy increases (decreases) when a quantum is subtracted from a super-Poissonian (sub-Poissonian) state. The subtraction does not influence a state governed by Poissonian statistics. In this work, however, we will exclusively analyze the subtraction from single-mode and multimode thermal states because we start from thermal equilibrium.

Furthermore, second-order correlation function $g^{2}(0)=\frac{\langle a^{\dagger 2}a^2\rangle}{\langle a^{\dagger}a \rangle^2}=
1+\frac{1}{1+m}$ for (\ref{subtr}) converges to unity, irrespectively to $n_{\text{th}}$. It is depicted in Fig.~\ref{fig:prms}(b). However, Fano factor $F=\langle (\Delta n)^2\rangle /\langle n\rangle=1+n_{\text{th}}$ is independent on $m$ and it approaches unity only for very small $n_{\text{th}}\ll 1$. To reach higher $\langle n\rangle$, $m$ needs to be higher too, which is increasingly more challenging to reach. Let us note that these results correspond to an instantaneous limit $\lambda t\rightarrow 0$ of the continuous photodetection process, where $\lambda$ is the success probability of single photon subtraction \cite{reff2}. For $n_{\text{th}}=2$, we experimentally demonstrate in Fig.~\ref{fig:prms}(c) that the conditional out-of-equilibrium statistics indeed remains super-Poissonian although $g^{(2)}(0)$ is substantially reduced below 2, which holds for thermal light. Figs.~\ref{fig:prms}(a,b,c) also show that the measured statistics and derived characteristics agree very well with the theoretical model.

Invariance of the Fano factor $F=1+n_{\text{th}}$ means that the variance $\langle (\Delta n)^2\rangle$ increases simultaneously with the increase of $\langle n\rangle$. However, $\langle (\Delta n)^2\rangle$ does not actually grow fast enough to render the state (\ref{subtr}) useless. For example, mean-to-standard-deviation ratio $\mbox{MDR}=\langle n\rangle/\sqrt{\langle (\Delta n)^2\rangle}=\sqrt{\frac{n_{\text{th}}}{1+n_{\text{th}}}}\sqrt{m+1}$ increases monotonously. It means, the energy advantageously increases faster than its fluctuations. Moreover, it is already sufficient to use $m=1$ and $n_{\text{th}}>1$ to obtain $\mbox{MDR}>1$ and the mean
$\langle n\rangle$ increases even faster with $m$ for larger $n_{\text{th}}$. Experimental evidence that $\langle n\rangle$ and $\mbox{MDR}$ increase with $m$ is shown in Figs.~\ref{fig:prms}(a,d). Using conditional instantaneous measurement, (\ref{subtr}) exhibits the same behavior of $\langle n\rangle$ and $\mbox{MDR}$ as thermal oscillator coherently driven out of equilibrium (see the Methods for details).

\begin{figure}
\centerline{\includegraphics[width=0.9\linewidth]{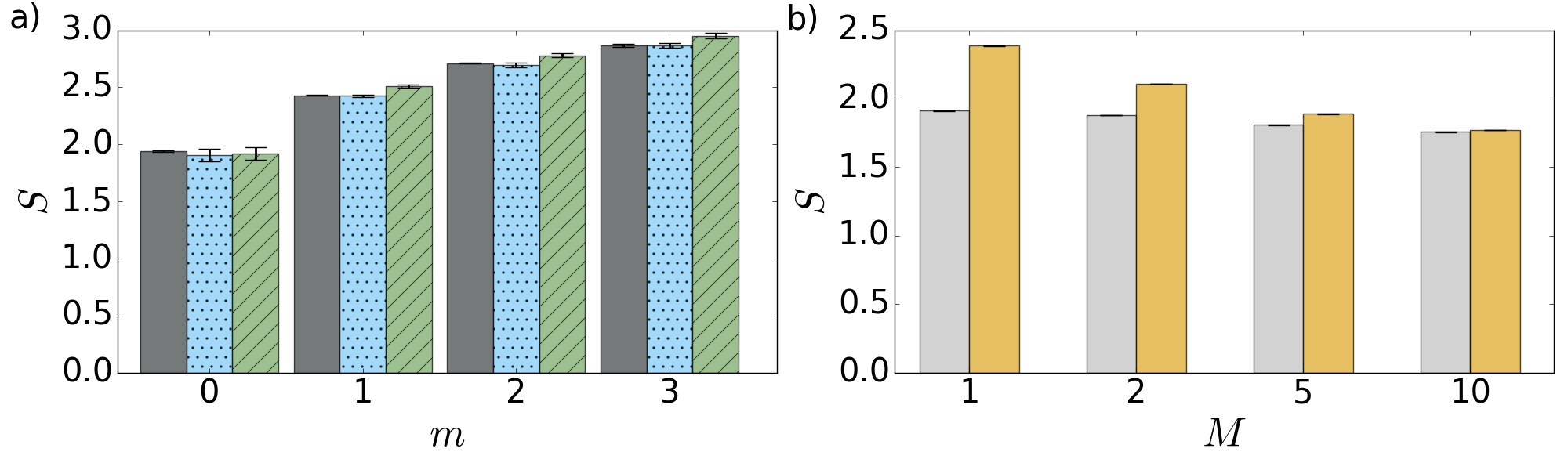}}
\caption{(a) Shannon entropy of the conditionally prepared states as a function of subtracted photon number $m$. Shown are experimental data (dark gray), full numerical model (blue dots), and the simplified model (green tiles) based on Eq.~(\ref{subtr}).
(b) Shannon entropy as a function of a number $M$ of modes of the initial thermal state. Gray bars stand for the entropy of $M$-mode thermal states ($m=0$) and yellow bars show the entropy of the same states after single-photon subtraction ($m=1$).}
\label{fig:S}
\end{figure}

\section*{Work available from out-of-equilibrium state}

Previous analysis suggests that the conditionally generated statistics (\ref{subtr}) can be a viable alternative to the oscillator externally driven out of thermal equilibrium. To support this statement, we predict and measure work available from the out-of-equilibrium state (\ref{subtr}). Available average work $\langle W\rangle_{\text{yield}}$, which is performed while the system equilibrates with the environment with temperature $T$, is expressed by relative entropy \cite{w1,w2,w3,w4,w6}
\begin{equation}\label{work}
\langle W\rangle_{\text{yield}}=k_BT D\left(p_n||p^{eq}_n\right).
\end{equation}
Here $k_B$ is Boltzmann constant and $D(p_n|p_n^{eq})=\sum_{n=0}^{\infty}p_n \ln p_n - \sum_{n=0}^{\infty} p_n \ln p_n^{eq}$ is relative Shannon entropy (Kullback-Leibler divergence) between the out-of-equilibrium statistics $p_n$ and the distribution $p^{eq}_n$ of a system in the equilibrium with an environment with temperature $T$ \cite{w1}. Differently from the previous statistical analysis, which takes into account only the system, $\langle W\rangle_{\text{yield}}$ depends on both the system state and the environment with constant temperature $T$. The bound (\ref{work}) can be reached; some specific protocols are already developed \cite{Mari}. 

Our preparation method actually uses two reservoirs, hot one ($T>0$) and cold auxiliary vacuum reservoir ($T=0$), see Fig.~\ref{fig:scheme}. However, the cold reservoir cannot be used to provide work (\ref{work}), without heating some of its modes up using an external source. We can, therefore, consider the hot reservoir at temperature $T>0$ and cool one mode to its ground state by a strong dissipation. The temperature $T>0$ of the thermal source is always constant in the experiment; consequently, the available work can be normalized by $k_B T$. By cooling the oscillator mode to the ground state, normalized work  $\frac{\langle W\rangle_{\text{yield}}}{k_BT}|_0=\ln\left[1+n_{\text{th}}\right]$ sets a benchmark for any useful conditional preparation of out-of-equilibrium state. If $\frac{\langle W\rangle_{\text{yield}}}{k_BT}|>\frac{\langle W\rangle_{\text{yield}}}{k_BT}|_0$, more work can be extracted from the state prepared conditionally by the measurement of a small part of dissipated energy (presented protocol) than by a complete cooling of one of the hot reservoir modes down.
The cooling to ground state can be challenging for many systems such as mechanical oscillators and, therefore, the conditional procedure may be preferable to achieve more work.

\begin{figure}[!t]
\centering
\includegraphics[width=0.95\linewidth]{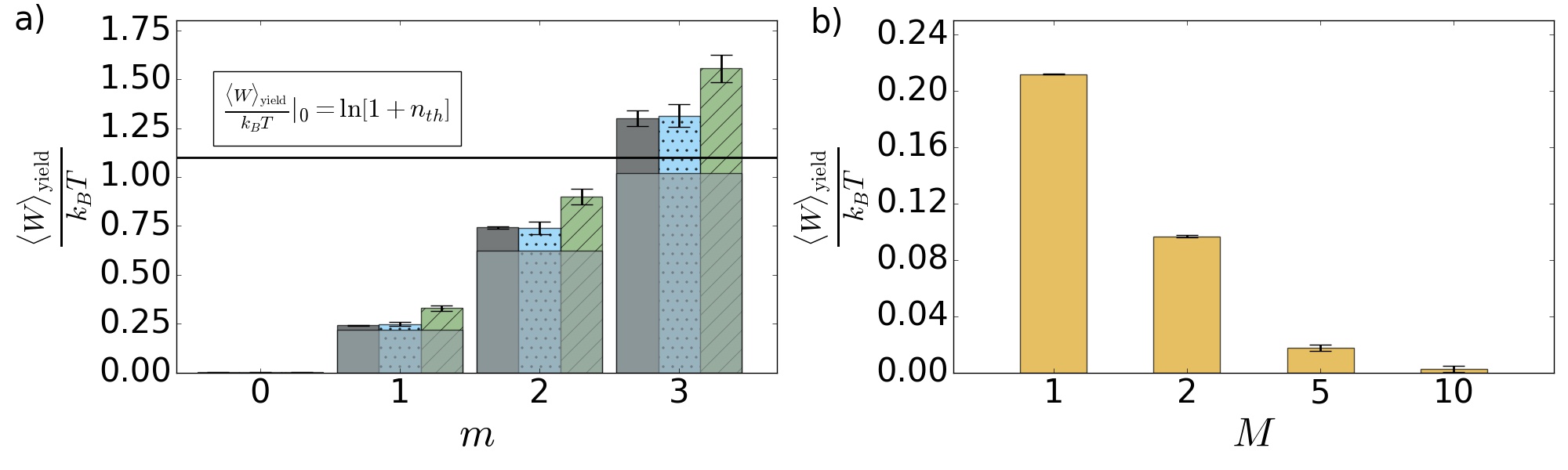}
\caption{(a) The normalized available work as a function of subtracted photon number $m$. Shown are experimental data (dark gray), full numerical model (blue dots), and the simplified model (green tiles) based on Eq.~(\ref{subtr}). The horizontal threshold (solid black line) corresponds to the work available by a cooling of the oscillator mode to the ground state. Light gray areas represent lower bounds derived for thermal state heated with the same mean number of photons as the corresponding $m$-photon subtracted states. (b) The normalized available work plotted against a number of modes $M$ for multimode thermal state after single-photon subtraction ($m=1$).}
\label{fig:W}
\end{figure}

For the statistics (\ref{subtr}), we conditionally obtain the normalized available work
\begin{equation}\label{work_th}
\frac{\langle W\rangle_{\text{yield}}}{k_BT}=\frac{1}{(1+n_{\text{th}})^{m+1}}\sum_{n=0}^{\infty}\left(\frac{n_{\text{th}}}{n_{\text{th}}+1}\right)^n \frac{(n+m)!}{n!m!}\ln\left[\frac{1}{(1+n_{\text{th}})^m}\frac{(n+m)!}{n!m!}\right],
\end{equation}
which increases monotonously with $m$ for any $n_{\text{th}}>0$ without an offset or saturation. We experimentally verified that for $n_{\text{th}}=2$, see Fig.~\ref{fig:W}(a). The entropy also increases with $m$, as shown in Fig.~\ref{fig:S}(a). The amount of extractable work decreases for increasing number $M$ of modes of the initial thermal state, see Fig.~\ref{fig:W}(b). It vanishes completely in the incoherent limit of large $M$. It clearly demonstrates that first-order coherence is a resource needed to extract available work using the instantaneous dissipation and photon measurement.

Despite the increase in entropy, the available work obtained by the subtraction procedure overcomes the threshold $\frac{\langle W\rangle_{\text{yield}}}{k_BT}|_0$ 
given by the complete cooling
already for $m=3$. The experimental results shown in Fig.~\ref{fig:W}(a) demonstrate the violation by 5 standard deviations. Moreover, the available work also overcomes a threshold set by a thermal state heated to the same mean number of quanta as reached by the subtraction. The work available by the adequate thermal heating is illustrated by a light gray area of the bars. Heating or cooling to the ground state -- heating/cooling strategy -- represents a joint benchmark here. All experimental results in Fig.~\ref{fig:W}(a) agree with the theory predictions. It opens the possibility to test other thermodynamical quantities and processes using the presented experimental photonic approach.

\section*{Information carried by out-of-equilibrium state}

\begin{figure}
\centering
\includegraphics[width=0.95\linewidth]{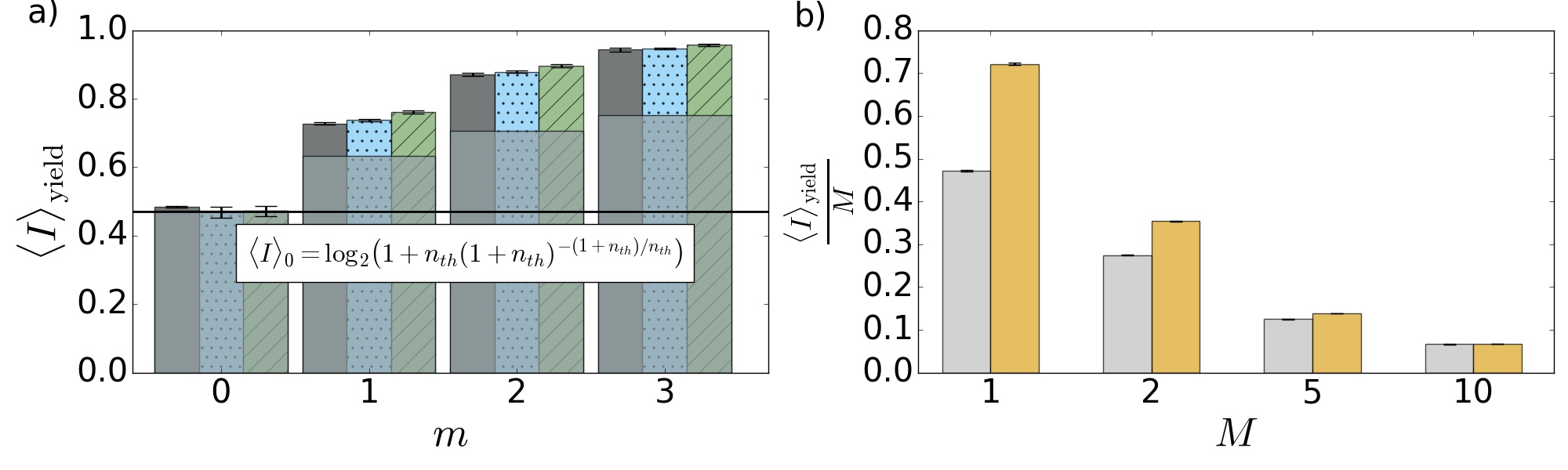}
\caption{(a) The maximum mutual information per mode against a number of subtracted photons $m$. Dark gray bars stand for experimental results, blue dots stand for full numerical model, and green tiles represent the ideal model based on Eq.~(\ref{subtr}). Light gray areas represent lower bounds derived for thermal state heated with the same mean number of photons as the corresponding $m$-photon subtracted states. Solid black line represents a threshold of maximum mutual information available when encoding `1' using the initial thermal state. (b) The maximum mutual information per mode as a function of the number of modes $M$ for multimode thermal state before and after $m$-photon subtraction. Colors refer to the value of $m$, gray: $m$=0, yellow: $m$=1.}
\label{fig:MI}
\end{figure}

We complement the measurement of available work by verification that the out-of-equilibrium distribution (\ref{subtr}), as a member of a binary alphabet, can carry information better than initial thermal distribution $p_{n,\text{th}}$. Average mutual information given in bits can be determined by the relative entropy
\begin{equation}\label{inf}
\langle I\rangle=D(p^{AB}_{i,j}||p^A_ip^B_j)
\end{equation}
where indices $i,j=0,1$ stand for a single bit at a sender side A and a single bit on a receiver side B, respectively. $p^{AB}_{i,j}$ is a joint (correlated) probability distribution and $p^A_i$, $p^B_j$ are marginal probability distribution at the sender and receiver sides, respectively. In contrast to the average work (\ref{work}), average information depends on joint statistics of both communicating parties. Also, it is optimal to use vacuum state of the cold  reservoir for encoding of the symbol `0'. The `1' can be encoded using thermal distribution $p_{n,\text{th}}$, which sets a benchmark $\langle I\rangle_{0}$ for the mutual information, see the Methods section for the details. To overcome the thermal bound, we employ the conditional statistics (\ref{subtr}) instead to encode the symbol `1'. In this case, the average mutual information reaches $\langle I\rangle=H\left((1-p^A_0)(1-p_E)\right)-(1-p^A_0)H(p_E)$, where $p_E=1/(1+n_{\text{th}})^{m+1}$ is the probability of error (symbol `1' is identified as `0') and $H(p_E)$ is a binary entropy function. The maximum
\begin{equation}
\langle I\rangle_{\text{yield}}=\mbox{max}_{p^A_0}\langle I\rangle_m=\log_2 \left(1+(1-p_E)p_E^{p_E/(1-p_E)}\right)
\end{equation}
of mutual information $\langle I\rangle$ over the probability $p^A_0$ at the sender side is monotonously increasing with $m$ for any $n_{\text{th}}$. The experimental result for $n_{\text{th}}=2$ is shown in Fig.~\ref{fig:MI}(a). It is not critically sensitive to a number of modes when a multimode thermal state is used instead of the single thermal mode. For the multimode states, the information gain has to be normalized per mode, because more modes can carry more information. It vanishes only gradually, as is presented in Fig.~\ref{fig:MI}(b). The first-order coherence is a key resource here, same as for the work extraction.

For arbitrary small $n_{\text{th}}$ and any $m>0$, average information $\langle I\rangle_{\text{yield}}\approx (1+m)n_{\text{th}}/(e\ln 2)$ overcomes the benchmark $\langle I\rangle_{0}$ for any $n_{\text{th}}$. The results of experimental verification for $n_{\text{th}}=2$ are shown in Fig.~\ref{fig:MI}(a). Information gain $\langle I\rangle_{\text{yield}}$ approaches its maximum of 1~bit even faster than for a thermal state heated to the equivalent mean number of quanta. We reach more than 0.9~bit already for $m=3$. Indeed, the conditionally generated state can carry maximum information despite its mixedness.

\section*{Conclusion}

We have experimentally produced the conditional out-of-equilibrium state (\ref{subtr}) from single-mode thermal light by a weak dissipation to a reservoir and an inefficient detection of photons there. We have theoretically and experimentally verified that average work can be extracted from the conditional out-of-equilibrium state, which outperforms any cooling/heating strategy. Furthermore, this state can also be used to carry more average information (closer to one bit) than for any state produced by a cooling/heating strategy, despite entropy increase of the conditional state. The presented procedure does not require any external coherent drive or additional thermal energy. It only uses energy measurement to reach higher work and information rate conditionally. However, it conclusively requires the first-order (classical) coherence of the thermal source.
Obtained results complement the previous experiments demonstrating the applications of the subtraction procedure.
The presented method can be translated to other experimental platforms and used for future experiments in currently merging fields of quantum information and quantum thermodynamics.
It is also stimulating for current optomechanical experiments at single quanta level \cite{marcus}, where a mechanical oscillator is driven out-of-equilibrium by a weak optical cooling and incoherent photon detection more efficiently than by a complete cooling or adequate heating.

\section*{Methods}

\subsection*{Experimental setup}

Subtraction of $m$ quanta from a thermal state was experimentally realized to demonstrate generation and characterization of out-of-equilibrium states of light.
The pseudothermal pulsed light was generated employing a nanosecond pulsed laser diode (805~nm) in gain switching regime with repetition rate of 4~MHz. This initial optical signal was focused on the surface of a rotating ground glass and the output speckle pattern was coupled into a single-mode optical fiber. The Glauber second order correlation function of the generated pseudothermal state was evaluated, $g^{2}\left(0\right)=2.00\left(3\right)$, to verify the high quality of the preparation stage.
Multimode thermal states were generated by selecting $M$ thermal modes with the same overall mean photon number $\langle n\rangle=n_{\text{th}}$ but different temporal modulation. The effective number of modes $M$ is modified by changing the size of speckle pattern collected via the optical fiber. This is achieved by changing either the diameter of the laser spot on the rotating ground glass or the distance between the glass and fiber coupler.

Multiple-photon subtraction was realized using low-reflectivity beam splitter implemented with a half-wave plate followed by a polarizing beam splitter.
In the first port, the reflected photons were detected via reconfigurable multichannel detector with $m$ commercial on-off single-photon detectors. To measure click statistics of the transmitted pulses, we placed photon-number-resolving detector (PNRD) at the second port. The PNRD consists of balanced eight-channel spatially multiplexed optical network and eight single-photon avalanche photodiodes. 
The resulting coincidence statistics was acquired by the PNRD under the condition that exactly $m$ detection events occurred at the reflected port.
We have applied a statistical method based on the maximum-likelihood algorithm to reconstruct resulting photon statistics from the multi-coincidence measurement.

The coincidence rates increase with increasing mean photon number of the initial thermal state. However, it is crucial to set the mean photon number low enough to measure a coincidence statistics of $m$-photon subtracted thermal state within the range of the PNRD. Mean photon number $n_{\text{th}}$ of the $m$-photon subtracted thermal state increases with $m$ by factor $(m+1)n_{\text{th}}$. Taking into account the number of channels of the PNRD and its efficiency, we can safely set $n_{\text{th}} = 2$ for the maximum number of subtracted photons $m\leq 3$. At the same time, the selected mean photon number is high enough to keep the measurement time reasonably short.

\begin{figure}
\centering
\includegraphics[width=0.7\linewidth]{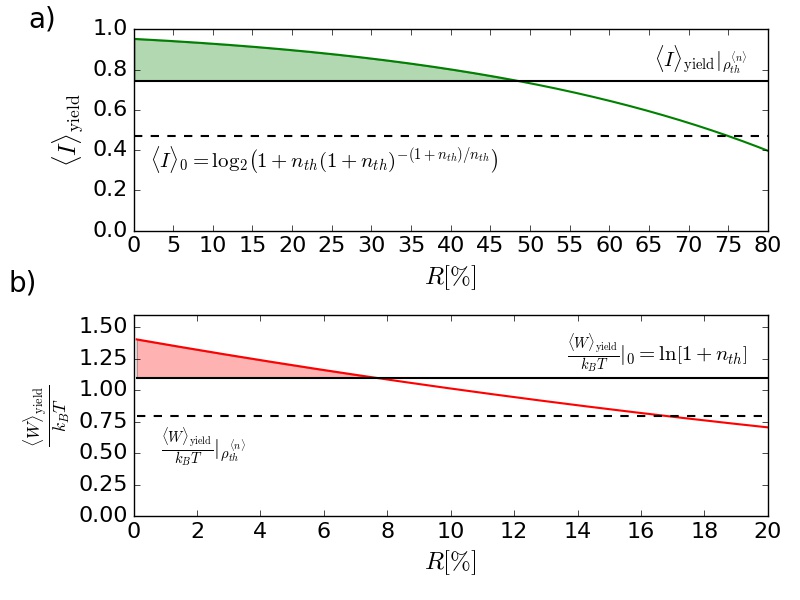}
\caption{(a) The maximum mutual information (solid green curve) and (b) the available work (solid red curve) versus the beam-splitter reflectivity $R$ for three-photon-subtracted thermal state ($m=3$). Solid black lines and dashed black lines represent the corresponding benchmarks discussed in the main text.}
\label{fig:R}
\end{figure}

Similarly, the value of beam-splitter reflectivity represents a trade-off between the subtraction rate (and, consequently, the total measurement time) and the ability of the generated out-of-equilibrium state to perform work and transfer information.
Both these quantities monotonously decrease with increasing reflectivity $R$ (see Fig.~\ref{fig:R} for the three-photon-subtracted state). We can see that the chosen value of the reflectivity, $R = 5\%$, is close to the maximum possible one, which outperforms the tightest bound on the available work.

To fully describe the performed $m$-photon subtraction, a detailed model has been developed. 
It takes into account actual experimental properties of the set-up: the beam-splitter reflectivity $R$, the number $m$ of on-off detectors at the reflected port, and their detection efficiency. In the limit of $R\rightarrow 0$, the full numerical model is equivalent to an application of $m$-th power of annihilation operator to the input thermal state, which produces the ideal statistics (\ref{subtr}). The numerical model has been used to evaluate all the parameters discussed in the main text and plotted in Figs.~\ref{fig:prms}-\ref{fig:MI}.

\subsection*{Out-of-equilibrium statistics vs. coherently driven thermal noise}

The out-of-equilibrium statistics (\ref{subtr}) is similar to a statistics of thermal oscillator coherently driven out of thermal equilibrium with mean
$\langle n\rangle_c=n_{\text{th}}+n_{c}$ and variance $\langle (\Delta n)^2\rangle_c = 2n_cn_{\text{th}}+n_c+n_{\text{th}}^2+n_{\text{th}}$, where $n_c$ is the mean number of coherent quanta caused by the driving. Considering $n_c=gn_{\text{th}}$, where $g$ is ratio between coherent and incoherent energy, we can see that both $\langle n\rangle_c$ and $\mbox{MDR}_c=\frac{\langle n\rangle_c}{\sqrt{\langle (\Delta n)^2\rangle_c}}$ monotonously increase with $g$ for any $n_{\text{th}}$, similarly to $\langle n\rangle_m$ and $\mbox{MDR}_m$ using the statistics (\ref{subtr}) in the case of $m$-quanta subtraction.

However, to reach $\mbox{MDR}_c>1$, $n_{\text{th}}<(n_c-1)n_c$ is necessary and, therefore, small coherent driving out of equilibrium is not sufficient for large $n_{\text{th}}$. For small $n_{\text{th}}\ll 1$, achievable only by cooling, the mean-to-deviation ratio reaches $\mbox{MDR}_c\approx \sqrt{n_{\text{th}}(1+g)}$, similarly as for $\mbox{MDR}_m$ with $m$ substituted by $g$. Simultaneously, the second-order correlation function
$g^{(2)}_c(0)=1+\frac{1}{1+\frac{g^2}{1+2g}}$ also does not depend on $n_{\text{th}}$ similarly as for $g^{(2)}_m(0)$, although it has different dependency on $g$. Fano factor $F_c=1+n_{\text{th}}+n_{\text{th}}\frac{1+2g}{1+g}$ depends on $g$, whereas $F_m$ is principally independent on $m$. However, for large $g$ it also does no converge to $F_c=1$ and Poissonian statistics. Only for small $n_{\text{th}}\ll 1$, both statistics converges to Poissonian limit. Despite (\ref{subtr}) is not statistics of thermal state coherently driven out of equilibrium, it exhibits similar statistical features without any coherent drive.

\subsection*{Benchmark for available work}

Let us evaluate the available work (\ref{work}) that is performed while the oscillator in an initial thermal state $p_{n,\text{th}}^{(1)}$ equilibrates with the environment in thermal state $p_{n,\text{th}}^{(2)}$ with temperature $T$. The oscillator retains its thermal Bose-Einstein statistics but the mean number of quanta $n_{\text{th}}^{(1)}$ decreases to $n_{\text{th}}^{(2)}<n_{\text{th}}^{(1)}$. For $n_{\text{th}}^{(2)}>0$, the normalized work reads
\begin{equation}\label{cool}
\frac{\langle W\rangle_{\text{yield}}}{k_BT}=D(p_{n,\text{th}}^{(1)}||p_{n,\text{th}}^{(2)})=n_{\text{th}}^{(1)}\ln\frac{n_{\text{th}}^{(1)}}{n_{\text{th}}^{(2)}}+(1+n_{\text{th}}^{(1)})\ln\frac{1+n_{\text{th}}^{(2)}}{1+n_{\text{th}}^{(1)}}.
\end{equation}
Expression (\ref{cool}) represents the lower bound (light gray areas) in Fig.~\ref{fig:W}(a). The mean number difference $\delta n_{\text{th}}=n_{\text{th}}^{(2)}-n_{\text{th}}^{(1)}<0$ corresponds to cooling of the oscillator, where thermal energy is dissipated to another reservoir at a lower temperature. Positive $\delta n_{\text{th}}>0$ means that the system has been heated, which requires additional source of thermal energy and therefore this case is not considered here.

\subsection*{Benchmark for carried information}

For two thermal distributions $p_{n,\text{th}}^{(1)}$ with mean number of photons $n_{\text{th}}^{(1)}$ (representing bit 1) and $p_{n,\text{th}}^{(0)}$ with $n_{\text{th}}^{(0)}<n_{\text{th}}^{(1)}$ (representing bit 0), the optimal measurement strategy distinguishes between number of quanta less or equal to $n_{\text{max}}$ (detection of bit 0) and higher than $n_{\text{max}}$ (detection of bit 1).
The maximum mutual information (\ref{inf}) over $p^A_0$ approaches
\begin{equation}\label{ass}
\mbox{max}_{p^A_0}I=\log_2\left(1+2^\frac{H(p_{01})-H(p_{10})}{1-p_{01}-p_{10}}\right)
-\frac{1-p_{10}}{1-p_{01}-p_{10}}H(p_{01})+\frac{p_{10}}{1-p_{01}-p_{10}}H(p_{10}),
\end{equation}
where and $H(x)=-x \log_2 x - (1-x)\log_2 (1-x)$ is binary entropy function, $p_{01}=1-\left(n_{\text{th}}^{(1)}/(1+n_{\text{th}}^{(1)})\right)^{1+n_{\text{max}}}$ is error probability of sending bit 1 and receiving it as bit 0, and $p_{10}=\left(n_{\text{th}}^{(0)}/(1+n_{\text{th}}^{(0)})\right)^{1+n_{\text{max}}}$ is error probability of sending bit 0 and receiving it as bit 1. To minimize the total error probability it is necessary to use two states whose distributions have the smallest possible overlap. For thermal states we can assume $n_{\text{th}}^{(0)}=0$, which yields $p_{10}=0$ and $n_{\text{max}}=0$. The optimal extraction of information is then simply the measurement of zero and non-zero energy. In this case, the maximum mutual information
\begin{equation}\label{bench}
\langle I\rangle_{0} =\mbox{max}_{p^A_0}I=
\log_2\left(1+n_{\text{th}}^{(1)}(1+n_{\text{th}}^{(1)})^{-(1+n_{\text{th}}^{(1)})/n_{\text{th}}^{(1)}}\right)
\end{equation}
monotonously increases with $n_{\text{th}}^{(1)}$, linearly as $n_{\text{th}}^{(1)}/(e\ln 2)$ for small $n_{\text{th}}^{(1)}$, and slowly saturates at 1 bit. The benchmark (\ref{bench}) sets a lower bound on mutual information available by using the vacuum state (bit 0) and a thermal state with the same mean photon number as the prepared $m$-photon subtracted state (bit 1). This bound is shown in Fig.\ref{fig:MI}(a) by light gray areas for individual $m$.

\section*{Acknowledgements}
This work was supported by the Czech Science Foundation (project GB14-36681G).
J.H. acknowledges the financial support of the project IGA-PrF-2017-008 of the Palacký University.
We thank Michal Dudka for the development of fast coincidence circuit used for multiphoton heralding and many discussions on electronic interface of the photon-number-resolving detector.

\section*{Author contributions statement}
J.H. constructed experimental setup, performed measurements and numerical simulations, and analyzed data. M.J. supervised the experiment and analyzed data. R.F. suggested the theoretical idea, performed calculations, and supervised the project. All authors participated in writing the manuscript.

\section*{Additional information}
The authors declare no competing financial interests. Correspondence and requests for material should be addressed to R.F. filip@optics.upol.cz.

\end{document}